# A Nanograins-attached and Ultrathin Cu Flake Powder Fabricated by High Energy Mechanical Milling and Dealloying

*Chenguang Li, Mingwei Zhang, Mianmian Ruan, Jun Wang, Jiamiao Liang\*, Deliang Zhang\**

Prof. Deliang Zhang
School of Materials Science and Engineering, Shanghai Jiao Tong University, Shanghai 200240, China
Shanghai Key Laboratory of Advanced High-temperature Materials and Precision Forming, Shanghai Jiao Tong University, Shanghai 200240, China
Key Laboratory for Anisotropy and Texture of Materials (Ministry of Education), School of Materials Science and Engineering, Northeastern University, Shenyang 110819, China
Key Laboratory of Data Analytics and Optimization for Smart Industry (Ministry of Education), Northeastern University, Shenyang 110819, China.

Dr. Jiamiao Liang
School of Materials Science and Engineering, Shanghai Jiao Tong University, Shanghai 200240, China
Shanghai Key Laboratory of Advanced High-temperature Materials and Precision Forming, Shanghai Jiao Tong University, Shanghai 200240, China
**E-mail:** zhangdeliang@sjtu.edu.cn; jmliang@sjtu.edu.cn

Dr. Chenguang Li, Dr. Mingwei Zhang, Mianmian Ruan, Prof. Jun Wang
School of Materials Science and Engineering, Shanghai Jiao Tong University, Shanghai 200240, China
Shanghai Key Laboratory of Advanced High-temperature Materials and Precision Forming, Shanghai Jiao Tong University, Shanghai 200240, China

**Keywords**: ultrathin Cu flake powder, Cu nanograin, high energy mechanical milling, dealloying

## Abstract

Metal powders with hierarchical nanostructures are always designed and fabricated by dealloying with or without combination of other manufacturing processes. However, they are mainly nanoporous metal powder and its derivations, and their monotonous nanostructures restrict their applications, so metal powders with novel nanostructures should be explored further for various applications. Herein, high energy mechanical milling and dealloying can be combined together for fabricating mass-produced metal powders with controllable nanostructures. As an example, a nanograins-attached and ultrathin Cu flake powder can be fabricated by high energy mechanical milling of a Cu-42wt.%Al powder mixture and subsequent dealloying. The dealloyed Cu powder particles have ultrathin flaky shapes with numerous Cu grains being attached to their surfaces, and the microstructure of the as-milled Cu-42wt.%Al powder particles and the dealloyed Cu particles is studied to elucidate the formation mechanism of the unique morphology of the dealloyed Cu powder.

It has been established that nanoporous metal structures can be obtained by dealloying of binary alloys such as Cu-Al,[1] Cu-Mn,[2] and Au-Ag.[3] The nanoporous structures can be applied in the field of catalysis, sensing and energy conversion/storage because of their high specific surface areas and hierarchical microstructure.[3] In the dealloying process, less noble atoms in the alloy are removed, and the more noble atoms in the alloy are left behind and self-organized to form crystals.[1, 4]. The microstructure of the dealloying products is usually controlled by adjusting the phase composition and microstructure of the precursor alloy and post-dealloying treatment.[3] There are three types of phase compositions of the precursor alloy: type I: one solid solution phase and one or more intermetallic phase(s); type II: dual intermetallic phases and type III: a single multicomponent phase.

High energy mechanical milling (HEMM) is widely used in fabrication of ultrafine grained and nanocrystalline metallic and metal matrix composite powders.[5] Metallic powders with all three types of phase compositions can be fabricated by adjusting powder composition and milling parameters. In this study, we explored a novel approach of fabricating a Cu powder by dealloying Cu-Al composite powder with a lamellar structure made by HEMM of a Cu-42wt.%Al powder mixture and found that Cu powder particles with an unique morphology of ultrathin flakes with numerous Cu nanograins attached on their surfaces formed. This paper is to report the results of the study and discuss the formation mechanism of the interesting dealloying product.

As shown in **Figure 1**(a) and (b), after 4h of milling of the Cu-42wt.%Al powder mixture, a lamellar Cu(Al)/Al(Cu) composite structure formed, with the thickness of Cu(Al) and Al(Cu) layers being at sub-micron scale. SEM-EDS Cu and Al mappings (**Figure 1**(c) and (d)) suggested that extensive interdiffusion between Cu and Al layers occurred during milling, causing formation of Al(Cu) and Cu(Al) solid solutions.

As shown in **Figure 3**, the XRD pattern of the as-milled powder didn't show any new phase other than the Al(Cu) and Cu(Al) phases, suggesting that the $Cu_xAl_y$ intermediate phases either did not form or their amount was too small to be detected by XRD. Furthermore, TEM bright field and dark field images and selected area electron diffraction (SAED) patterns (**Figure 2**(a)-(c)) revealed the existence of $CuAl_2$ and $Cu_9Al_4$ phases in the as-milled powder. The content of Al and Cu (**Figure 2**(d)) at Sp1 to Sp3 in **Figure 2**(a) as determined by EDS point analysis showed positive concentration gradient of Cu and negative concentration gradient of Al from the Al(Cu) layer to Cu(Al) layer. Based on this analysis, we can conclude that Al(Cu) and Cu(Al) solid solutions formed by interdiffusion between Al and Cu, and the Al(Cu)/Cu(Al) interfaces have $CuAl_2$ and $Cu_9Al_4$ intermetallic layers. The thickness of Al(Cu) and Cu(Al) layers and the sizes of $CuAl_2$ and $Cu_9Al_4$ layers were all in nano-scale.

As shown by the XRD pattern of the dealloyed powder in **Figure 3**, after dealloying, the Al peaks disappeared from the XRD pattern, which indicates that most of Al atoms were removed by dealloying. Meanwhile $Cu_2O$ peak appeared because of surface oxidation of the dealloyed powder particles by acidic solution used in dealloying and subsequent washing, filtration, drying processes. Composition analysis of the Cu flaky powder using ICP showed that the Al and Fe contents of the dealloyed powder were 0.25wt.% and 0.05wt.% respectively, which means the purity of the dealloyed powder was higher than 99.5%.

As shown in **Figure 4**, the dealloyed Cu powder particles had flaky shapes and there were numerous polyhedral nanograins being attached on their surfaces (**Figure 4**(a) and (b)). Black spots with a polygonal shape were observed in the TEM bright field image (**Figure 4**(c)) of the dealloyed Cu particles supported by the Cu grid of TEM specimen. They were thickness contrasts caused by the nanograins attached to the surfaces of the Cu powder particles. The Cu flake powder particle substrates were polycrystalline with grain sizes in the sub-micron scale. The surface morphology of the dealloyed Cu flake powder particles were further examined by AFM (**Figure 4**(b)), and it revealed that the powder particle surfaces not only had Cu nanograins attached to them, but also showed islands/channels structures around the attached Cu nanograins. This means that the ultrathin Cu flakes have a hierarchical surface morphology. Based on the above observation, the dealloyed Cu powder is named as nanograins attached ultrathin Cu flake powder (the NGA powder). The specific surface area of the NGA powder was measured by BET to be 1.964$m^2$/g, which was around 5-9 times of that of the Cu flake powder with flat particle surfaces and comparable thickness.[6]

It has been established that HEMM of a Al/Cu powder mixture can cause formation of Al(Cu)/Cu(Al) lamellar composite structure with intermetallic particles distributed at the Al(Cu)/Cu(Al) interfaces,[7] so it is not necessary to discuss the formation mechanism of the structure of the as-milled Cu-42wt.%Al alloy powder particles during HEMM. The dealloying process of the Cu-Al alloy powder involves removing Al atoms and self-organization of the residual Cu atoms to form crystals.[4a, b].The morphology and structure of the Cu powder particles formed by dealloying are closely related to the phase composition of the precursor powder particles. When there are multiple phases in the composite powder, they tend to be dealloyed orderly and dealloying behavior is different for different phases. Liu *et al* [8] investigated the dealloying behavior of an Al-15at.%Cu alloy consisting of Al(Cu) and $CuAl_2$ phases, and revealed that the Al(Cu) phase was dealloyed preferentially because of its higher corrosion potential and lower corrosion current comparing to $CuAl_2$. After dealloying of the Al(Cu) phase, numerous autogenous Cu nanograins formed on the surface of the remaining $CuAl_2$ phase. At the second stage of dealloying, the $CuAl_2$ phase was dealloyed to form nanoporous structure. This sequential dealloying caused by the multiple phases in Al-Cu alloy and their corresponding dealloying microstructure shed light on the formation mechanism of the NGA powder during dealloying of the as-milled Cu-42wt.%Al composite powder in the present study.

As illustrated in the schematic diagrams shown in **Figure 5**, the formation of the NGA powder particles can be divided into three stages. In stage I, pure Al is etched away by the acid solution accompanying with the disassembly of the Cu-Al lamellate structure and the exposure of Al(Cu) solid solution to the acid solution. In stage II, dealloying of the Al(Cu) solid solution occurs, and causes the formation of Cu nanograins which are attached to surface of intermetallic particles and the Cu(Al) flakes.[8] Finally in stage III, the $CuAl_2$ and $Cu_9Al_4$ phases are dealloyed, leading to formation of an islands/channels structure on the surfaces of the remaining Cu(Al) flakes. The Cu(Al) solid solution may be hard to be dealloyed, since its Al content may be lower than the critical content of Al needed for dealloying.[4b] In summary, after the etching of pure Al and the sequential dealloying of Al(Cu) solid solution and $CuAl_2$ and $Cu_9Al_4$ phases, the unique surface morphology composing of numerous nanograins being attached to the Cu flake surfaces which also have a more subtle islands/channels structure form.

This kind of ultrafine grained and ultrathin Cu flake powder with Cu nanograins attached to the flake surfaces has a great potential to be used to fabricate a bulk Cu material with a unique microstructure consisting of ultrafine grained and ultrathin Cu laminates and individual nanometer sized interlamellar Cu nanograins by powder consolidation. Such bulk Cu materials may have a good combination of both high strength and high ductility.[9] Furthermore, the large specific surface area of the NGA powder may make it advantageous in being used as catalysts and sensors with high efficiency and stability.[10]

An ultrathin and ultrafine grained Cu flake powder with numerous Cu nanograins attached to the flaky powder particle surfaces was successfully fabricated by a combination of HEMM and subsequent dealloying. The ultrafine and ultrafine grained Cu flakes were the corresponding Cu(Al) layers in the lamellar Al(Cu)/Cu(Al) composite structure formed by HEMM of Cu-42wt.% powder mixture, while the Cu nanograins attached to the Cu flake surfaces and the islands/channels structure of Cu flake surfaces were obtained by dealloying of Al(Cu) phase and $Cu_xAl_y$ intermetallic layers distributed at Al(Cu)/Cu(Al) interfaces. The NGA powder with such a special flake surface morphology and microstructure has a potential for being applied to fabrication of advanced Cu materials by powder metallurgy, catalysis and sensing.

**Experimental Section**

*HEMM and dealloying processes:* A Cu-42wt.%Al alloy powder was prepared by HEMM of a mixture of a Cu powder with a dendritic particle shape (purity≥99.9%, 45μm, Shanghai ST-nano Science and Technology Co., Ltd, China) and an Al powder with a spherical particle shape (purity≥99.9%, 45μm, Henan Yuanyang Powder Technology Co., Ltd, China). The scanning electron microscopy (SEM) secondary electron images of the as-received powder particles are shown in **Figure 6**. The milling process was conducted on a QM-3SP4 planetary ball mill (Nanjing Nanda Instrument Ltd.). 0.7wt.% steric acid together with elemental Cu and Al powders were sealed in a vial within a glove box filled with argon to avoid powder welding to the wall of the vial and oxidation. Prior to HEMM, a low speed (200rpm) milling was carried out for 4h to mix the powders. Then the mixed powder was milled for 4h at a high speed of 500rpm. The ball to powder weight ratio was 10:1. The as-milled powder was dealloyed using a dilute hydrochloric acid (1.8 mol/L) to remove Al atoms from the powder. The powder/acid solution mixture was filtered with deionized water by suction filtration, and the solid powder was vacuum dried for 6h at 110°C.

*Characterization:* The phase composition and microstructure of the as-milled and dealloyed powders were characterized by X-ray diffractometry (XRD) with a scanning step of 0.02° and a speed of 10°/min, scanning electron microscopy (SEM) (Mira 3 and Sirion 200) together with energy dispersive spectroscopy (EDS) (Aztec X-act), transmission electron microscope (TEM) (JEM2100F) and atomic force microscopy (AFM) (Dimension Icon & FastScan Bio). The impurity content and the specific surface area of the dealloyed powder was measured using inductively coupled plasma emission spectrometer (ICP) (ICAP 6000 Radial) and BET method (ASAP 2460), respectively.

**Acknowledgements**

This work was supported by the National Natural Science Foundation of China (No.51271115).

**References**

[1] Liu, W. B., Zhang, S. C., Li, N., Zheng, J. W., Xing, Y. L., *Corros. Sci.* **2011,** *53* (2), 809.
[2] Hayes, J. R., Hodge, A. M., Biener, J., Hamza, A. V., Sieradzki, K., *J. Mater. Res.* **2006,** *21* (10), 2611.
[3] Song, T., Yan, M., Qian, M., *Corros. Sci.* **2018,** *134*, 78.
[4] a) Erlebacher, J., Seshadri, R., *MRS Bull.* **2011,** *34* (8), 561; b) McCue, I., Benn, E., Gaskey, B., Erlebacher, J., *Annu. Rev. Mater. Res.* **2016,** *46* (1), 263; c) Zhao, C., Qi, Z., Wang, X., Zhang, Z., *Corros. Sci.* **2009,** *51* (9), 2120; d) Zhang, Z. H., Wang, Y., Qi, Z., Somsen, C., Wang, X. G., Zhao, C. C., *J. Mater. Chem.* **2009,** *19* (33), 6042; e) Zhang, Z. H., Wang, Y., Qi, Z., Zhang, W. H., Qin, J. Y., Frenzel, J., *J. Phys. Chem. C* **2009,** *113* (29), 12629; f) Lu, H. B., Li, Y., Wang, F. H., *Scr. Mater.* **2007,** *56* (2), 165.
[5] a) Zhang, D. L., *Prog. Mater Sci.* **2004,** *49* (3-4), 537; b) Murty, B. S., Ranganathan, S., *Int. Mater. Rev.* **1998,** *43* (3), 101; c) Suryanarayana, C., *Prog. Mater Sci.* **2001,** *46* (1-2), 1; d) Suryanarayana, C., Al-Aqeeli, N., *Prog. Mater Sci.* **2013,** *58* (4), 383.
[6] Tan, Z., Li, Z., Fan, G., Li, W., Liu, Q., Zhang, W., Zhang, D., *Nanotechnology* **2011,** *22* (22), 225603.
[7] Zhang, D. L., Ying, D. Y., *Mater. Sci. Eng., A* **2001,** *301* (1), 90.
[8] Liu, W. B., Zhang, S. C., Li, N., Zheng, J. W., An, S. S., Xing, Y. L., *Int. J. Electrochem. Sci.* **2012,** *7* (3), 2240.
[9] a) Wu, H., Fan, G. H., Huang, M., Geng, L., Cui, X. P., Xie, H. L., *Int. J. Plast.* **2017,** *89*, 96; b) Xiong, D. B., Cao, M., Guo, Q., Tan, Z., Fan, G., Li, Z., Zhang, D., *Sci. Rep.* **2016,** *6*, 33801.
[10] a) Sun, S. N., Li, H. Y., Xu, Z. C. J., *Joule* **2018,** *2* (6), 1024; b) Chen, C., Qi, Z. M., *Opt. Mater. Express* **2016,** *6* (5), 1561.

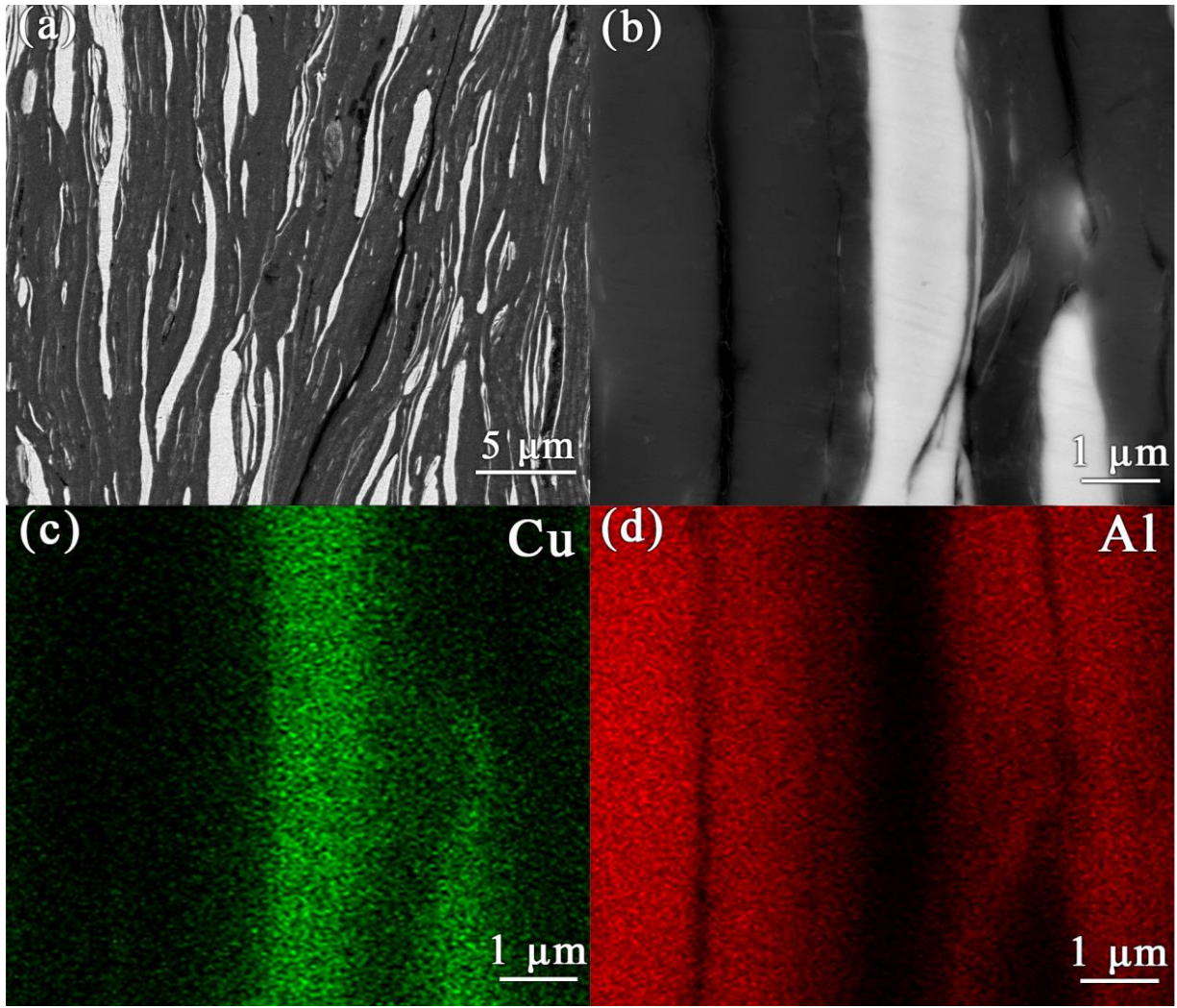

**Figure 1.** (a) and (b) SEM back scattered electron images of the cross section morphology and (c) and (d) corresponding EDS elemental maps of Cu and Al of an as-milled powder particle.

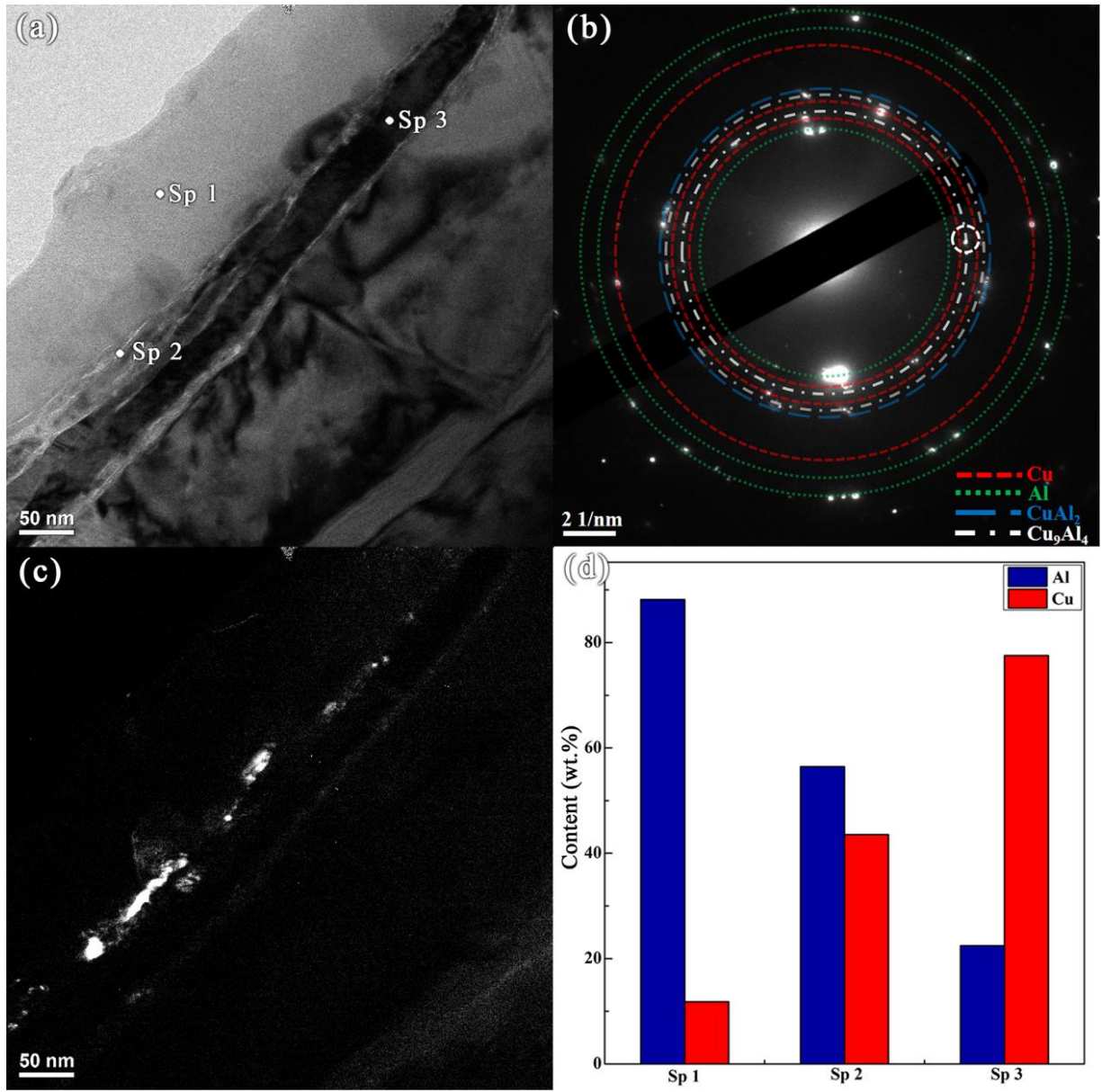

**Figure 2.** (a) TEM bright field image, (b) the SAED pattern of the central region in (a), (c) dark field image formed using the diffraction spot which is marked with a white circle in (b), and (d) the content of Al and Cu at Sp 1 to Sp 3 of the cross section of an as-milled powder particle.

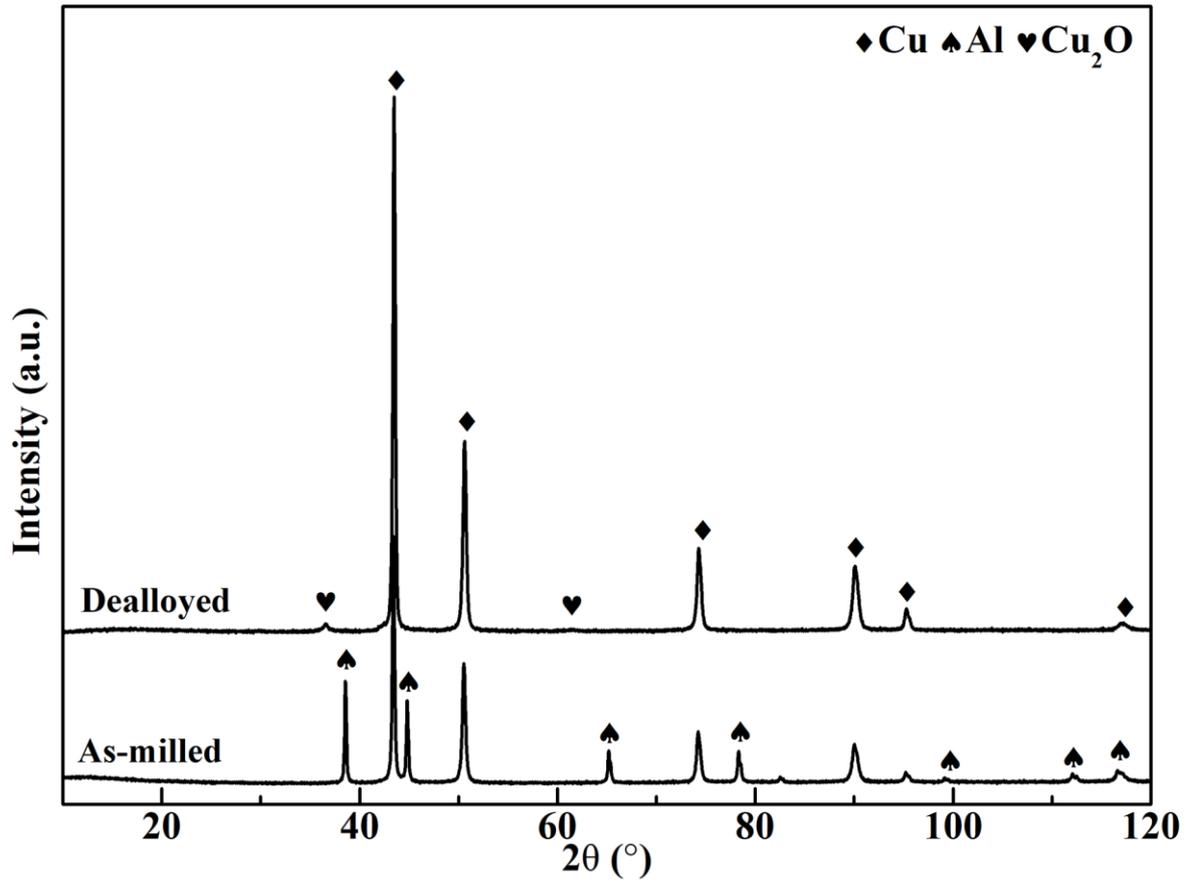

**Figure 3.** XRD patterns of the as-milled and dealloyed powders respectively.

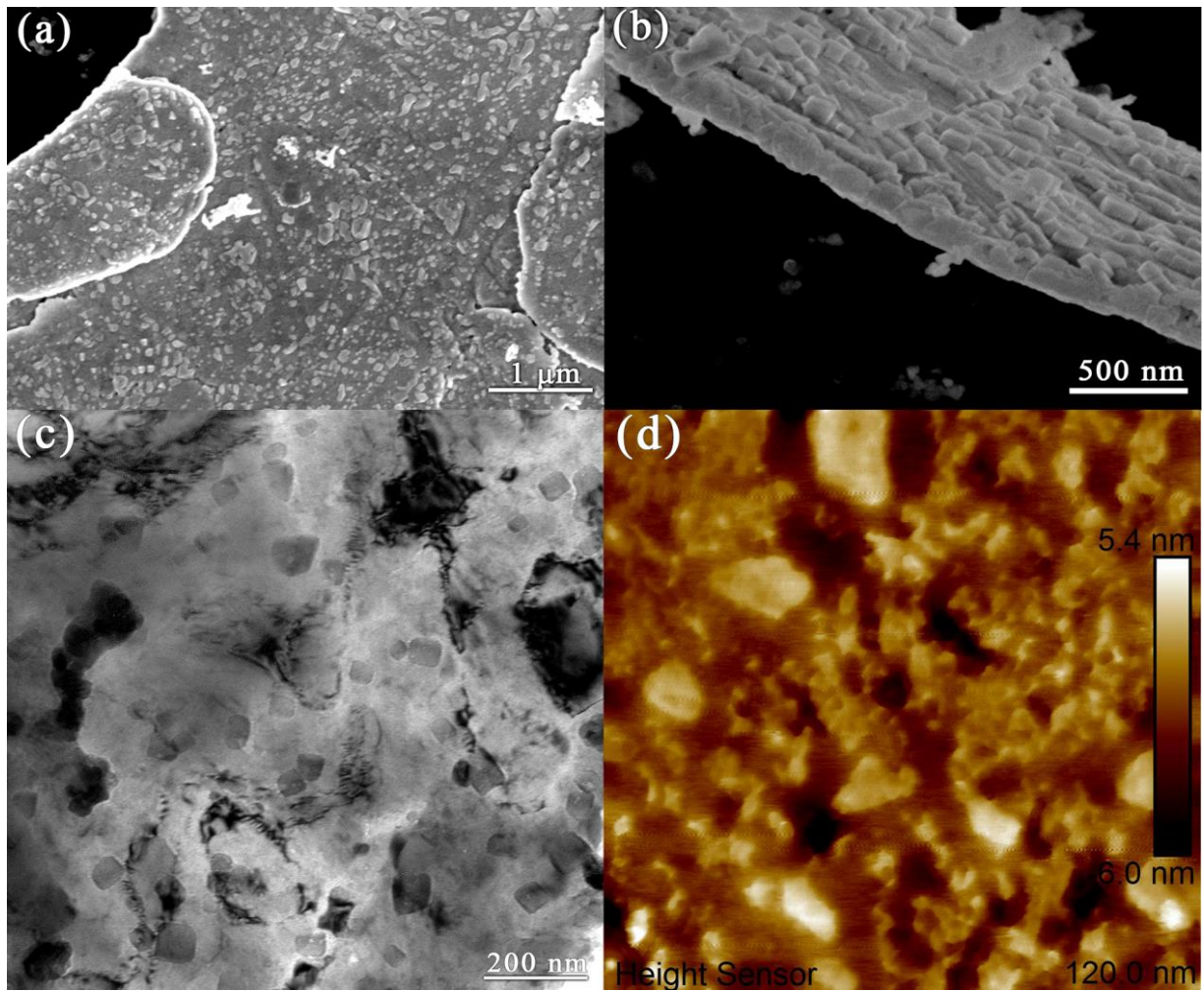

**Figure 4.** (a) and (b) SEM secondary electron images, (c) TEM bright field image and (d) AFM image of dealloyed powder particles.

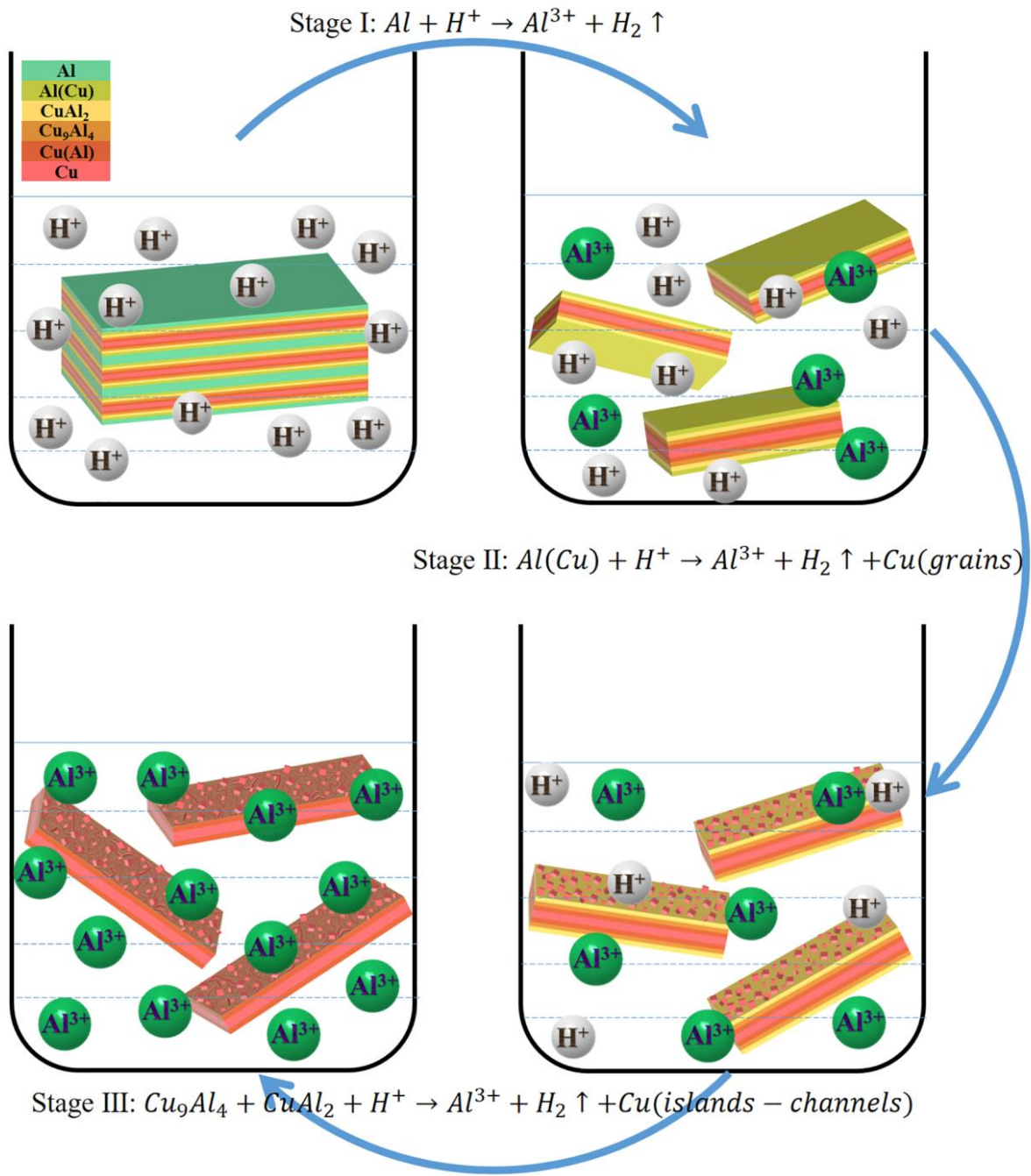

**Figure 5.** Schematic diagrams showing the formation process of the NGA powder.

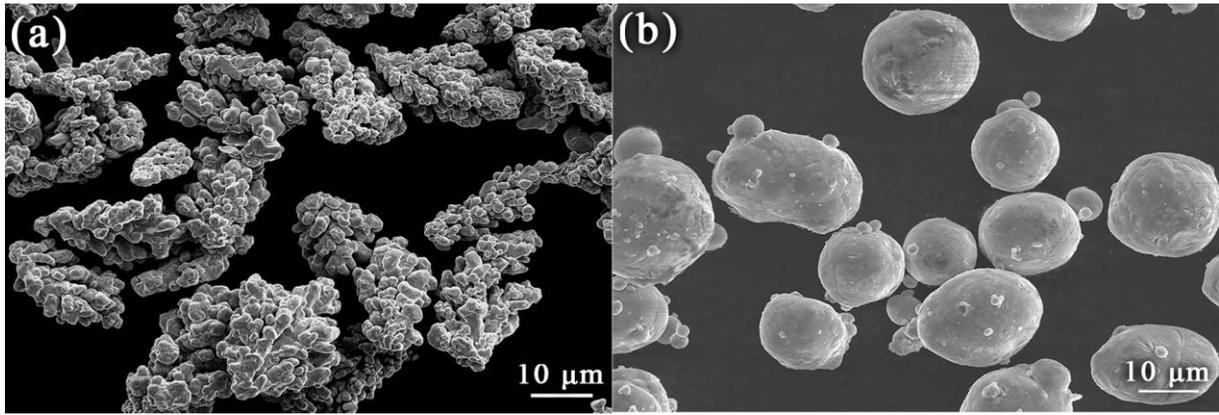

**Figure 6.** SEM secondary electron images showing the shapes and surface morphologies of the as-received (a) Cu and (b) Al powder particles.